\documentclass[preprint,12pt]{elsarticle}

\usepackage{amssymb}
\usepackage{color}
\usepackage{url}

\usepackage[T1]{fontenc}
\usepackage[utf8]{inputenc}

\journal{Chemical Physics}

\begin{document}

\begin{frontmatter}



\title{O$_2$ on Ag(110): A puzzle for exchange-correlation functionals}


\author[inst1]{Ivor Lon\v{c}ari\'{c}}
\author[inst2,inst3]{Maite Alducin}
\author[inst2,inst3,inst4]{J. I\~{n}aki Juaristi}

\affiliation[inst1]{organization={Ru{\dj}er Bo\v{s}kovi\'{c} Institute},
            addressline={Bijeni\v{c}ka 54}, 
            city={Zagreb},
            postcode={10000}, 
            country={Croatia}
            }

\affiliation[inst2]{organization={Centro de F\'{i}sica de Materiales CFM/MPC (CSIC-UPV/EHU)},
            addressline={Paseo Manuel de Lardizabal 5}, 
            city={Donostia-San Sebasti\'{a}n},
            postcode={20018}, 
            country={Spain}}
            
\affiliation[inst3]{organization={Donostia International Physics Center (DIPC)},
            addressline={Paseo Manuel de Lardizabal 4}, 
            city={Donostia-San Sebasti\'{a}n},
            postcode={20018}, 
            country={Spain}}

\affiliation[inst4]{organization={Departamento de Pol\'{i}meros y Materiales Avanzados: F\'{i}sica, Qu\'{i}mica y Tecnolog\'{i}a, Facultad de Qu\'{i}micas (UPV/EHU)},
            addressline={Apartado 1072}, 
            city={Donostia-San Sebasti\'{a}n},
            postcode={20080}, 
            country={Spain}}

\begin{abstract}
Despite the great success of density functional theory in describing materials, there are still a few examples where current exchange-correlation functionals fail. We add another example to this list that drives further development of functionals. We show that the interaction of O$_2$ with Ag(110) cannot be properly described by some of the most popular GGA, meta GGA, and hybrid functionals. We identify problems and provide clues for a functional that should be able to describe this and similar systems properly. 
\end{abstract}

\begin{graphicalabstract}
\includegraphics{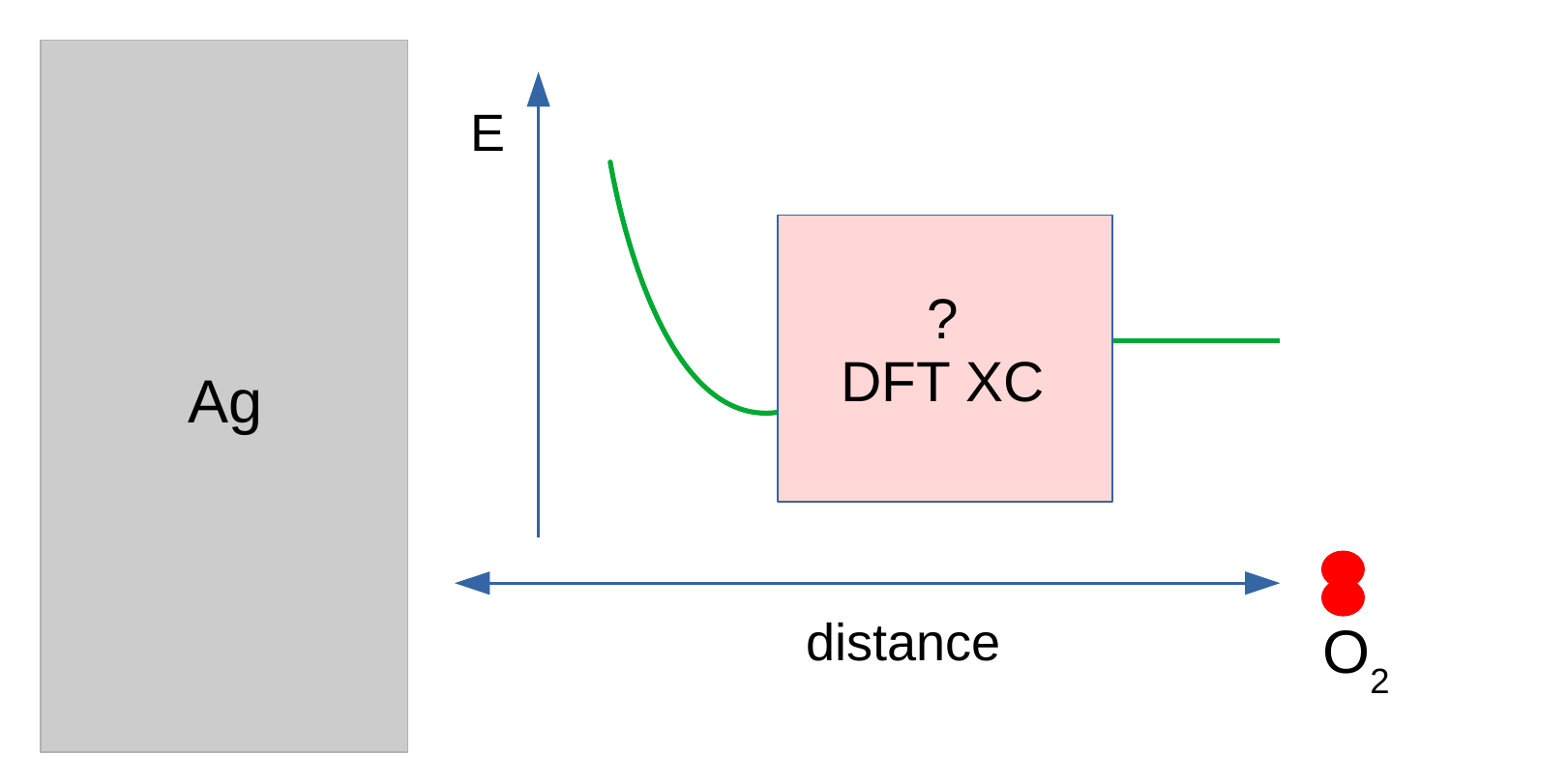}
\end{graphicalabstract}

\begin{highlights}
\item Interaction of O$_2$ with Ag(110) is challenging to describe with common exchange-correlation functionals
\item Hybrid meta-GGA functionals could provide needed accuracy
\end{highlights}


\end{frontmatter}


\section{Introduction}
\label{sec:intro}
The goal of theoretical and computational catalysis is to be able to model catalytic processes under operando conditions. To achieve this, an accurate potential energy surface (PES) that governs the reactions is necessary. Ideally, the method to obtain PES should be accurate and computationally manageable. Due to its good ratio of accuracy and computational complexity, density functional theory (DFT) is the most used method. Its accuracy and to some degree computational complexity depends on the exchange-correlation (XC) functional.

Since there is a lack of good benchmarking methods for the typical catalytic scenario of a molecule on a surface, mainly due to system size, DFT was often benchmarked to experimental data. DFT predictions can be compared to different experimental methods. From methods that provide information on structural properties, such as adsorption sites, to methods that give access to the energetics of interaction such as temperature-programmed desorption.

Particularly important are experiments that probe a large part of the configurational space of the PES. Additionally, experiments should be clean so that they can be easily compared to theoretical calculations. One such experimental technique is molecular beam experiments in high vacuum with well-defined surface face and initial conditions of molecules. Probabilities for different adsorption scenarios as well as analysis of scattered molecules are then possible \cite{sitz02, kleyn2003, juurlink09, vattuone10, beck2016}. On the other hand, using pulse probe experiments, often in laser-induced reactions on surfaces, the interaction time-scale can be also obtained \cite{Frischkorn2006}.

The theoretical description of molecular beam experiments requires a proper characterization of the system dynamics as well as an accurate evaluation of the gas/surface interaction~\cite{Gross98,C2CP42471A,kroescsr2016,Jiang2016,Jiang2019}
If theoretical calculations can accurately predict the results of such experiments, then the theoretical PES can be considered accurate. In many cases, DFT-based PES can nicely reproduce molecular beam experiments (see for instance, Refs. \cite{C2CP42471A,PhysRevLett.97.056102,Diaz06112009,hundt14,PhysRevLett108096101,petuya15,Nattino2016a,Loncaric2017,NOURGHASSEMI2017329,penatorres18,Rivero21}) and femtosecond laser desorption experiments~\cite{PhysRevLett.123.246802,Serrano2021,Scholz2016, Scholz2019}.

In other cases, the calculated PES is not accurate, which 
encourages the development of better theoretical methods, often in form of better XC functionals for DFT. 
Famous examples include the so-called CO adsorption puzzle in which DFT incorrectly predicted the adsorption site of CO on some metal surfaces \cite{Feibelman2001,PhysRevLett.98.176103,Olsen2003}. The problem was solved by taking care of dispersive forces that are not treated well with common XC functionals \cite{Hsing2019, schimka2010accurate, PhysRevB.81.045401,Loncaric2019}. Also remarkable are the efforts  to achieve chemical accuracy in the DFT based PES to describe chemical reactions at surfaces (see Ref.~\cite{kroes2021} for a recent review on this topic).

In previous studies, we constructed 6D PES of O$_2$ on Ag(110) using the most used XC functional in solid-state systems, the so-called PBE functional \cite{Perdew1996}. Using this PES, as well as full-dimensional ab-initio molecular dynamics, we modeled existing molecular beam experiments \cite{Loncaric2015, Loncaric2016}. Several discrepancies were found. The most important is that experiments found low adsorption probability at low incidence energies, whereas PBE-based simulations predicted high adsorption probability at low incidence energies. Also, in experiments, most of the adsorption is in the so-called chemisorption well which is not predicted with PBE based dynamics.

With the conclusion that standard PBE functional does not model O$_2$ on Ag(110) well, in this article we calculate some important PES features with different XC functionals and discuss their results in relation to experiments. We identify the problems of current XC functionals and give new perspectives for solving this puzzle.

\section{Methods}
\label{sec:methods}

All DFT calculations are performed with the \emph{Vienna Ab initio Simulation Package} (VASP)~\cite{VASP1,VASP2} (version 5.4.1). 
The Ag(110) surface is represented by a supercell of a five-layer slab with 14 layers of vacuum and a (2$\times$3) surface unit cell.
We apply the projector augmented-wave method (PAW)~\cite{PAW1, PAW2} to describe the electron--core interaction and use a cut-off of 400~eV in the plane-wave basis set.
The Brillouin-zone of the supercell is sampled with a $4\times4\times1$ $\Gamma$-centered Monkhorst-Pack grid of special $\mathbf{k}$-points~\cite{PhysRevB.13.5188}.
Partial occupancies are determined by the Methfessel-Paxton method~\cite{PhysRevB.40.3616} of order 1 with a smearing width of 0.2~eV. 

Different XC functionals are used in this study, as discussed below. A nice systematization of XC functionals is the so-called Jacob's ladder of DFT~\cite{jacobdft}. On the way to ``DFT heaven'' of accurate functional, the first step, and the simplest approximation, is the local density approximation, which is not considered here due to its well-known deficiencies for gas-surface systems.

The next step is the generalized gradient approximation (GGA) that exists in many forms. Excellent numerical stability and good accuracy make GGA functionals the most popular choice in solid systems. As already mentioned, PBE 
falls in this category, and it is popular for its overall good accuracy both for solids and molecules. PBE often overbinds molecules both in vacuum, resulting in too large atomization energies, and on transition metal surfaces, resulting in too large adsorption energies. 

The RPBE func\-tion\-al~\cite{RPBE} with modified dependence of the exchange energy on the normalized density gradient corrects this problem. However, this worsens the description of the solids and for other gas-surface systems, RPBE gives too repulsive PESs~\cite{n2w_xc,n2w_xc2,Wijzenbroek2014}. It has often been observed that experimental results for adsorption on transition metals lie in between the results obtained by PBE and RPBE and that their mixing can provide chemical accuracy for some systems~\cite{Diaz06112009, C2CP42471A} (see also Ref.~\cite{kroes2021} for a recent review on the applicability of this method and improved versions to different systems). 
Furthermore, both PBE and RPBE predict too large lattice constants. In this respect, PBEsol \cite{PBEsol} provides a better description of solids by revising some of the parameters used in constructing the PBE functional.

A major deficiency of typical semi-local GGA XC functionals is the lack of nonlocal dispersive interactions. Nowadays, many XC functionals that treat van der Waals forces exist. To see the effect of including these interactions we add to PBE the Tkatchenko-Scheffler correction~\cite{PhysRevLett.102.073005} as implemented in VASP~\cite{Bucko2016}.

The next step on the ladder are meta GGA functionals and in this study we used the following flavors: TPSS~\cite{TPSS}, RTPSS~\cite{Perdew2009}, MS0~\cite{Sun2012}, MS1~\cite{Sun2013}, MS2~\cite{Sun2013}, and M06L~\cite{Zhao2006}. The TPSS and RTPSS functionals are constructed in a similar way to the PBE and PBEsol functionals to match exact quantum-mechanical constraints without empirical parameters. They can at the same time describe molecules and solids better than PBE and PBEsol. The MS0 functional features a simpler functional form than (R)TPSS, containing further constraints, and shows better predictions for simple molecules and solids than (R)TPSS. Its name comes from ``made simple''-MS and ``zero''-0 fitting parameters. MS1 and MS2 are functionals with one and two fitting parameters, respectively, which were fitted to a database of several measurements and high-quality quantum chemistry calculations. Finally, M06L is a  functional with a large number of parameters fitted to a broad database. Meta-GGAs are in general numerically trickier to achieve self-consistency. For this reason, and to speed up calculations, we have used PBE geometries in evaluating energy as a function of molecule-surface distance.

A step above on the ladder, are the so-called hybrid functionals that contain a portion of exact exchange. In this study we used two screened hybrid functionals, HSE06~\cite{hse,hse06} and HSEsol~\cite{hsesol} and one meta hybrid functional MS2h~\cite{Sun2013}. HSE06 (HSEsol) is based on the PBE (PBEsol) functional and constructed such that 25\% of the exchange at short range is replaced by the exact exchange.  At long range, this functional is identical to the respective GGA functional. The good performance of these functionals for both solids and small molecules is well established~\cite{hsevasp,hsesol}. Finally, in MS2h 9\% of the MS2 exchange is replaced by the exact exchange. For all hybrid functional calculations we used PBE geometries as well. Since meta hybrid functionals are not implemented in the code version we used, our MS2h calculations are non-self-consistent. Energies reported here are obtained from self-consistent MS2 calculations, and afterwards 9\% of the MS2 exchange is replaced by the exact exchange energy calculated for the MS2 orbitals.

\section{Results}
\label{sec:results}

According to PBE, the PES of O$_2$ on Ag(110) is characterized by four adsorption wells \cite{roy,Loncaric2015, Loncaric2016}. In the four cases, the molecule lies parallel to the surface. There are two wells in the hollow site of the surface with the  molecule oriented 
in the $<1\bar{1}0>$ and <001> directions. One well is on the short bridge site in the  $<1\bar{1}0>$ direction, and the last one is on the long bridge site in the <001> direction. Adsorption energies are similar in all wells, and depending on the study, range from 0.2 to 0.4 eV \cite{Loncaric2016}. Adsorption in hollow sites is characterized by smaller surface-molecule distance, and larger Oxygen-Oxygen distance, and therefore, these wells are usually associated with the experimentally measured chemisorption wells. On the other hand, adsorption in bridge sites is characterized by larger surface-molecule distance, and smaller Oxygen-Oxygen distance, and therefore, these wells were associated with the experimentally measured physisorption wells. According to PBE, all wells are accessible from vacuum without a barrier. The configurational space to reach bridge wells is considerably larger than to reach hollow wells.

As a result, molecular dynamics on PBE PES predict barrierless adsorption, primarily to bridge wells from where the barrier to dissociation is large. This is in stark contrast to experiments that predict: (i) a noticeable barrier to adsorption and (ii) adsorption and dissociation in hollow wells. In this study, we calculate and study representative parts of the PES with different XC functionals. Due to the fact that, on the one hand, there is a great similarity between both hollow wells and that, on the other hand, this is also the case between both bridge wells, we focus only on the hollow and short-bridge wells in which the molecule axis is parallel to the $<1\bar{1}0>$ direction.

As a representative part of the PES, we calculate the adsorption energy of the molecule as a function of the distance from the surface for optimized O-O distance. These one-dimensional (1D) cuts of the PESs provide information on both adsorption energy and distance from the surface as well as the existence of any entrance barrier. Because the O$_2$ adsorption dynamics is rather direct~\cite{Loncaric2015, Loncaric2016}, this constitutes in principle the most important information from which one can infer the suitability of the PES to explain experimental results.

\begin{figure}[ht!]
    \centering
    \includegraphics{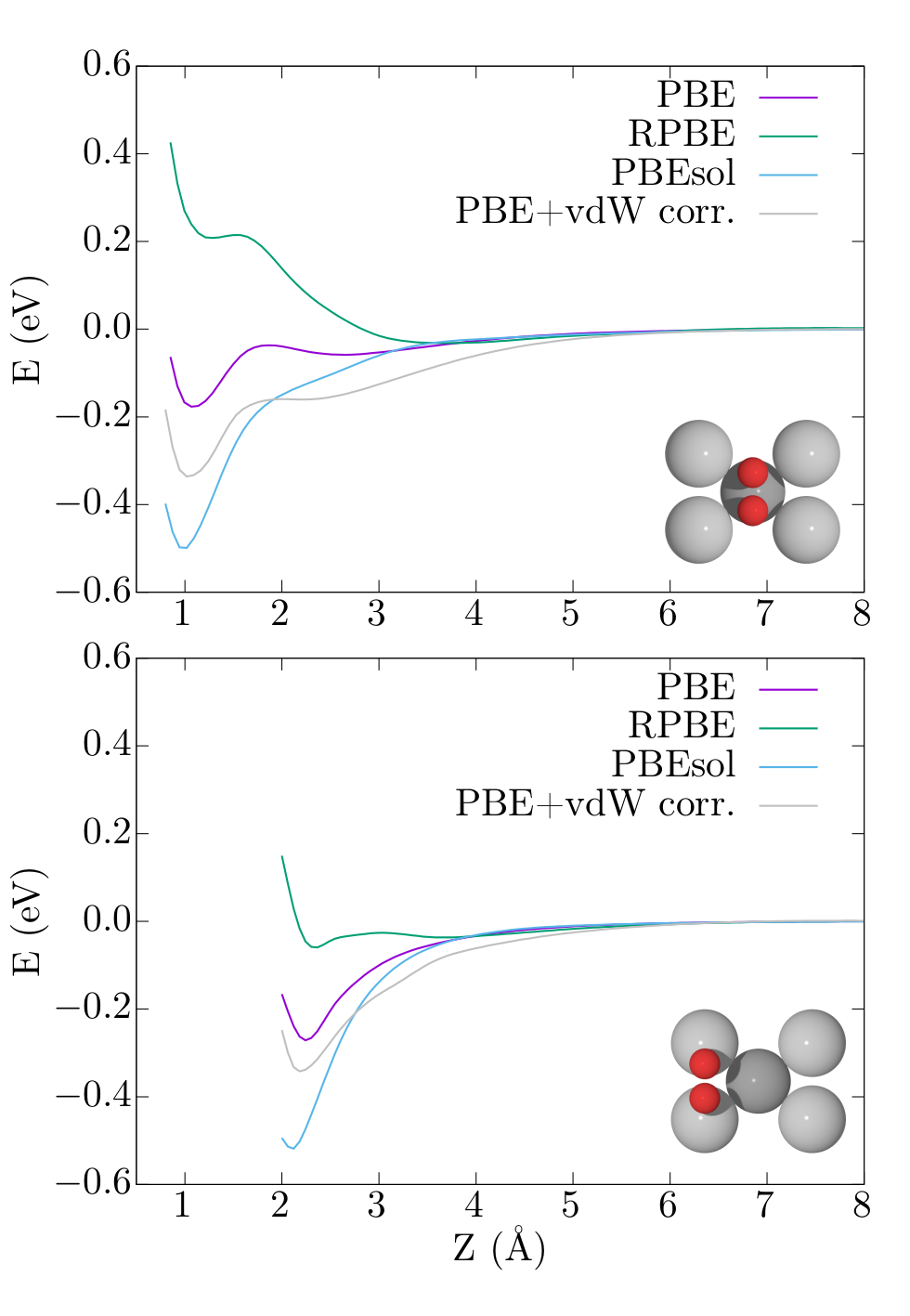}
    \caption{Adsorption energy of O$_2$ as a function of distance from the Ag(110) surface above hollow site (upper panel) and short bridge site (lower panel) calculated with different GGA functionals.}
    \label{fig:gga}
\end{figure}

In Fig. \ref{fig:gga} we show 1D PESs for the hollow and short bridge site calculated with different GGA functionals.
As already discussed above, the adsorption well at the hollow site is closer to the surface (at around 1 \AA) than the adsorption well at the short bridge site (somewhat larger than 2 \AA). While the position of the well is similar in all GGA functionals, there are large differences in adsorption energy. As expected, RPBE is the most repulsive functional while PBEsol is the most attractive, with PBE having intermediate adsorption energies. Clearly, there is no barrier for adsorption in the short bridge well with any of the functionals, while RPBE even predicts that the hollow well is not stable in respect to the free molecule. This means that common GGA functionals cannot reconcile the experimental fact of low adsorption probability at low incidence energies in molecular beam experiments. Moreover, as in the PBE case, it seems that other functionals would also predict most of the adsorption in bridge wells, while experiments suggest adsorption in hollow wells. Correction of PBE for van der Waals interactions lowers adsorption energies. The change is larger for the hollow well. While the van der Waals correction does not solve any of mentioned problems, it is worth pointing out that van der Waals corrections would increase adsorption in hollow wells compared to bridge wells in accordance with experiments. Let us also mention, that for this system, it is clear that mixing different GGA functionals~\cite{Diaz06112009} would not enable an accurate description of the adsorption process.

\begin{figure}[ht!]
    \centering
    \includegraphics{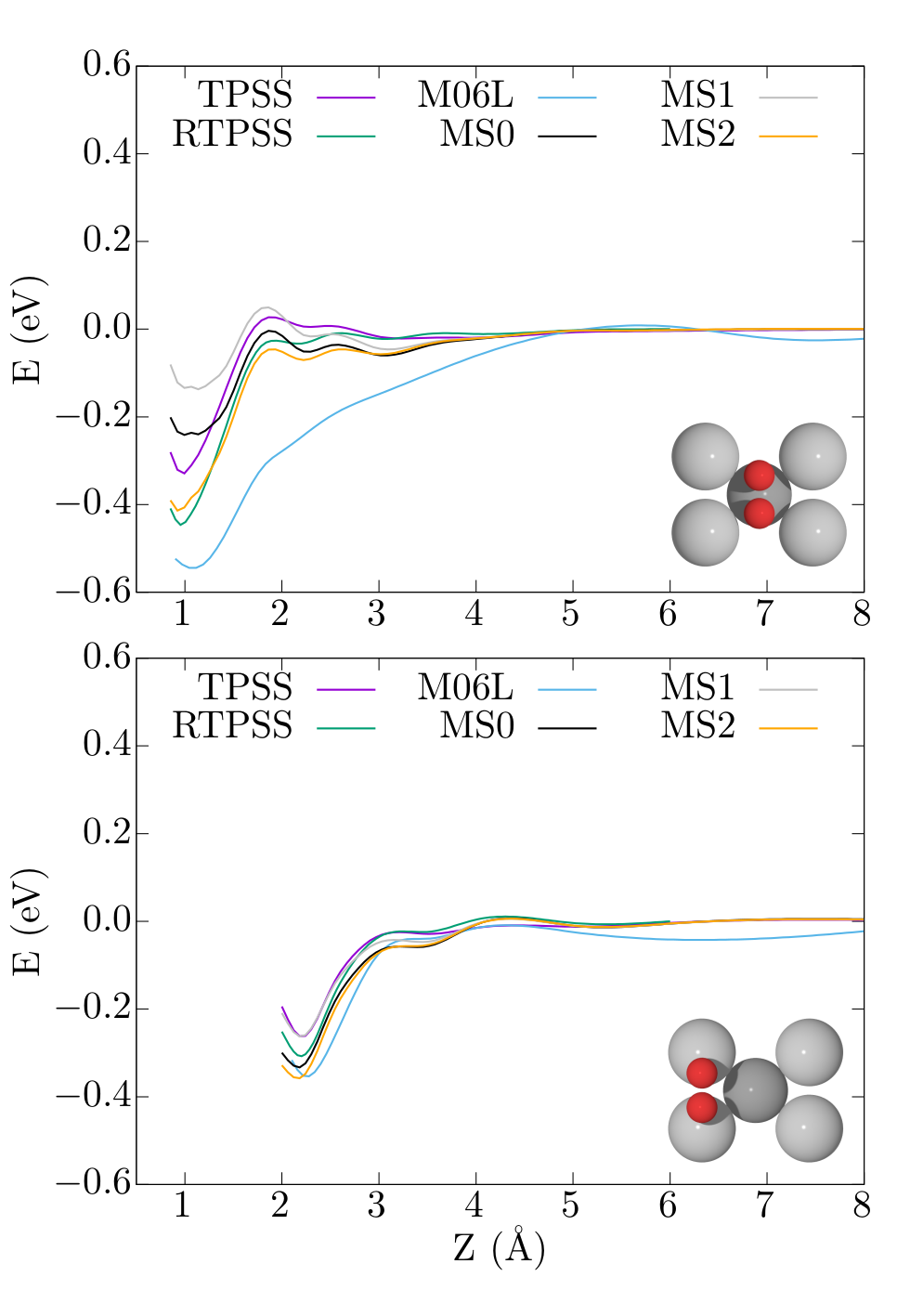}
    \caption{Adsorption energy of O$_2$ as a function of distance from the Ag(110) surface above hollow site (upper panel) and short bridge site (lower panel) calculated with different meta GGA functionals.}
    \label{fig:mgga}
\end{figure}

Fig. \ref{fig:mgga} shows results for meta GGA XC functionals.
Most meta GGAs predict similar energies in the entrance part of the PES. The only exception is M06L that predicts a strong attraction over the hollow well.  This result is somewhat reasonable considering that, in contrast to the other meta GGA functionals,  M06L is known to also capture dispersive interactions. 

Second, while all meta GGA functionals predict the adsorption energy in the short bridge well to be in a similar interval of 0.25-0.35 eV, the adsorption energy varies significantly for the hollow well. Values range from 0.13 eV for MS1 to 0.54 eV for  M06L. Four out of the six studied functionals, namely, 
M06L, TPSS, RTPSS, and MS2, predict the hollow well to be more stable than the short bridge well. In this respect, it seems that meta GGA could provide at least a partial solution to the O$_2$/Ag(110) puzzle. Still, while adsorption energies in hollow wells could be associated to the measured chemisorption state, adsorption energies in the short bridge well seem too large to be associated to the measured physisorption state.

\begin{figure}[ht!]
    \centering
    \includegraphics{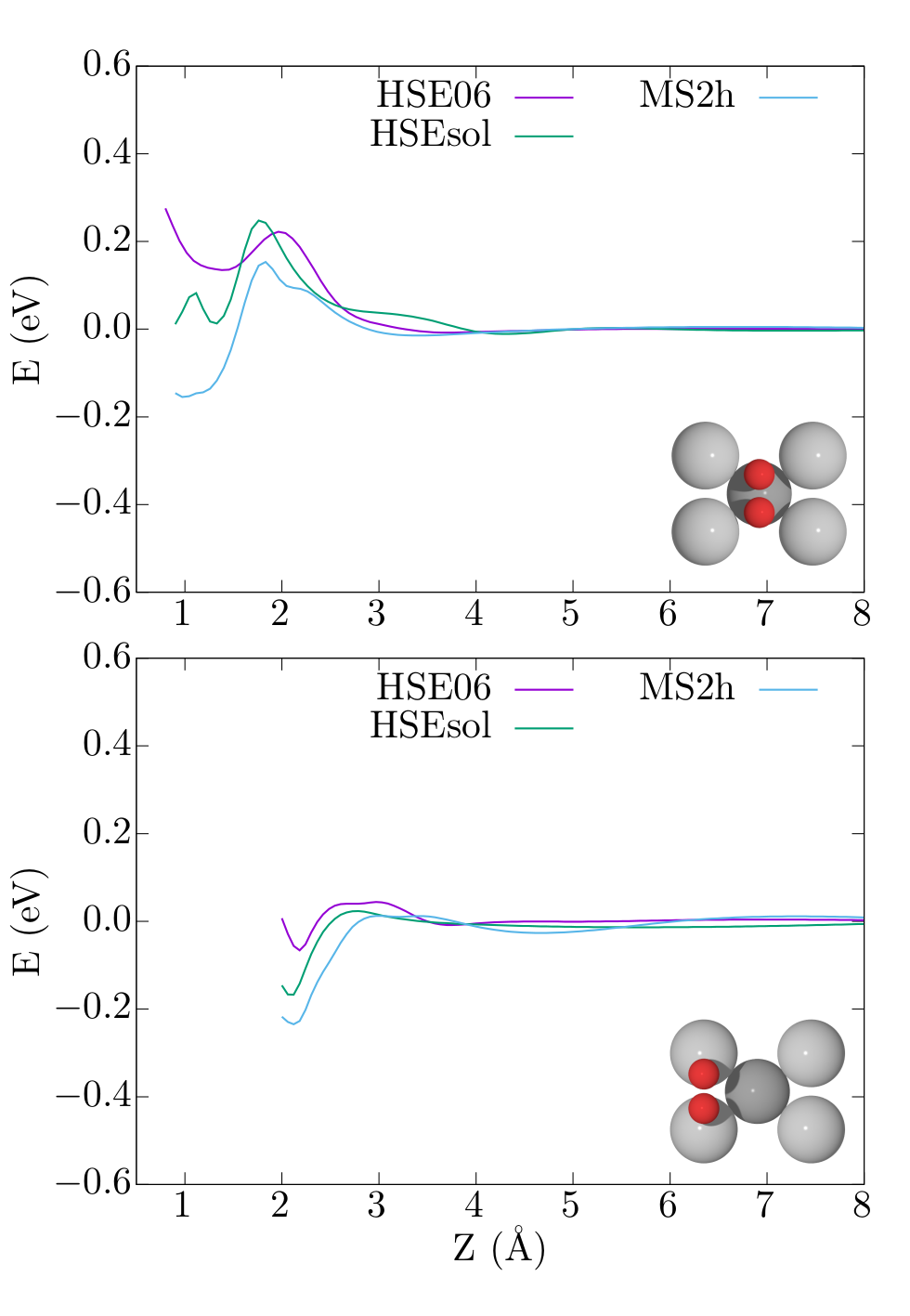}
    \caption{Adsorption energy of O$_2$ as a function of distance from the Ag(110) surface above hollow site (upper panel) and short bridge site (lower panel) calculated with different hybrid functionals.}
    \label{fig:hybrid}
\end{figure}

Fig. \ref{fig:hybrid} shows results for hybrid XC functionals.
Hybrid functionals are more repulsive in the entrance channel, showing a noticeable barrier of more than 0.1 eV at the hollow site and a small barrier of as little as 0.01 eV at the short bridge site. In this respect, hybrid functionals might be able to resolve the first disagreement between DFT and experiment concerning the measured low sticking probability at low incidence energies. However, all studied hybrid XC functionals predict larger adsorption energies in the short bridge well compared to the hollow well. Thus, they are again in contradiction with experiments. Since both HSE06 and HSEsol even predict that the hollow well would be unstable compared to vacuum, MS2h seems to be the best choice. It should be emphasized again that energies are evaluated at PBE interatomic distances, and it is possible that the adsorption energy could somewhat change with proper relaxation. Nevertheless, we do not expect that it would significantly change the conclusions.

\begin{figure}[ht!]
    \centering
    \includegraphics{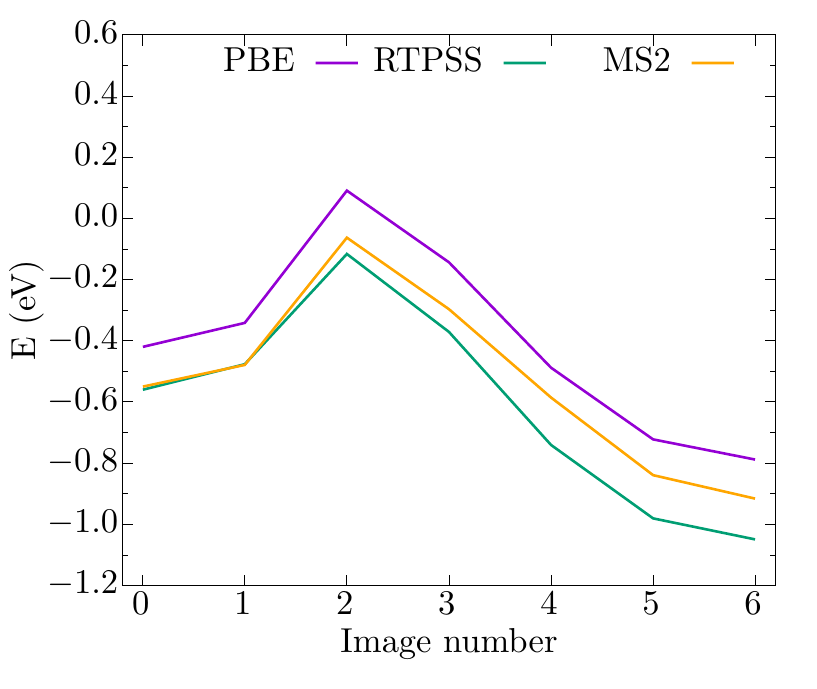}
    \caption{Minimum energy path for dissociation from the hollow well calculated with CI-NEB method with PBE, RTPSS, and MS2 XC functionals.}
    \label{fig:diss}
\end{figure}

Another difficulty with the PBE predictions is that the energy required to desorb the molecule is smaller than the energy required to dissociate it, in contrast to experiments~\cite{Vattuone1994_eels,rocca,rocca2,kleyn,campbell}. Therefore, we have also calculated the dissociation energy barriers shown in Fig.~\ref{fig:diss}, using the climbing image nudged elastic band method (CI-NEB) \cite{neb1,neb2,neb3,neb4} with the same computational setup as in Ref.~\cite{Loncaric2016}.
In contrast to PBE, the meta GGA functionals RTPSS and MS2 predict that it is more favorable to dissociate than to desorb, in accordance with experiments. Actually, the difference between the desorbing and dissociating energy barriers of around 0.1 eV agrees with the experimental value that is obtained from the decrease of sticking probability with surface temperature. 

\section{Disscusion and conclusions}
\label{sec:conclusions}
In this study, we have used different XC functionals and tested their predictions for the important PES features of O$_2$ on Ag(110). From GGA to meta GGA and hybrid XC functionals it seems that none of the tested functionals could reproduce the experimental results. Still, this study provides important guidelines on how to find the appropriate XC functional. It seems that the problem of low sticking at low incidence energies could be resolved using the hybrid functionals that tend to predict the existence of entrance barriers. Hybrid functionals should be based on a functional that predicts larger adsorption energy in hollow wells than in bridge wells. As shown here, most of the meta GGA functionals are successful in this respect. We have also shown that the inclusion of van der Waals forces stabilizes hollow wells compared to bridge wells. Along these lines, it should be possible to calculate the O$_2$/Ag(110) PES that would reproduce experimental results.

Problems in the DFT description of Oxygen interaction with metals are not unique to Ag(110). Regarding other Silver surfaces, for Ag(100) one finds very similar problems in reconciling DFT and experiments \cite{maite100}. Ag(111) is a much more closed surface without significant molecular adsorption, so that the problems are masked in this case. As a consequence, dissociation and scattering dynamics can be successfully modeled already at the GGA level \cite{itziar111,PRLitziar}. Another well-known case is the lack of entrance barrier for the interaction of O$_2$ with Al(111) \cite{PhysRevB.52.14954, PhysRevLett.92.176104, PhysRevLett.84.705}. While different approaches for the resolution of this problem have been suggested, for example, spin selection rules \cite{PhysRevLett.94.036104, PhysRevB.77.115421}, the problem lies in the ability of GGA XC functionals to properly model the interaction \cite{PhysRevLett.109.198303,yin2018}. Only with hybrid XC functionals, it is possible to obtain the barrier in the entrance channel using DFT \cite{hybrid_liu}. Problems in obtaining entrance barriers are likely the same for both Aluminum and Silver surfaces, supporting the conclusion that a portion of the exact exchange is needed to obtain accurate PESs.

Even though many experiments for O$_2$ on Ag(110) have been performed, for the final solution of the current problem, it is very important to perform additional precise reference experiments. In particular, due to the lack of benchmarking theoretical methods, it would be advantageous to have adsorption energies reliably measured, which can be achieved by, \textit{e.g.},  microcalorimetry~\cite{Fiorin2010}. The physisorption state is also not studied in detail by existing molecular beam experiments, and therefore such experiments at low temperatures would be advantageous. Also, due to different adsorption wells existing in this system, laser-induced dynamics could provide plenty of information \cite{Loncaric2016a, Loncaric2016b}.

From the theoretical point of view, it is desirable to try even more advanced methods than the here studied DFT XC functionals. Among  possible methods, it is worth highlighting many-body perturbation theory for the XC contribution \cite{schimka2010accurate}, embedded correlated wavefunction methods \cite{PhysRevLett.109.198303}, and quantum Monte Carlo methods \cite{Hsing2019}.
Each of these methods is numerically significantly more challenging than the DFT functionals used here and with its own advantages and disadvantages. It is questionable whether the full PES could be constructed using these methods, but at least they could be used as a benchmarking method.

All in all, we have presented a puzzle for DFT in describing O$_2$ on Ag(110). It seems that different GGA, meta GGA, and hybrid DFT XC functionals are not able to solve this puzzle. We have provided clues for a functional that could make it. We hope that this study will prompt further studies, both from the theoretical and experimental point of view along our suggestions.

\section{Acknowledgments}
This work has been supported in part by Croatian Science Foundation under the project UIP-2020-02-5675. We acknowledge financial support by the Gobierno Vasco-UPV/EHU Project No. IT1246-19 and the Spanish Ministerio de Ciencia e Innovación [Grant No. PID2019-107396GB-I00/AEI/10.13039/501100011033].



 \bibliographystyle{elsarticle-num} 
 \bibliography{cas-refs}





\end{document}